# Tree-based Single LED Indoor Visible Light Positioning Technique

Srivathsan Chakaravarthi Narasimman
*School of Electrical and Electronics Engineering,*
Nanyang Technological University, Singapore
srivaths003@e.ntu.edu.sg

Arokiaswami Alphones
*School of Electrical and Electronics Engineering,*
Nanyang Technological University, Singapore
EAlphones@ntu.edu.sg

*Abstract*— Visible light positioning (VLP) has gained prominence as a highly accurate indoor positioning technique. Few techniques consider the practical limitations of implementing VLP systems for indoor positioning. These limitations range from having a single LED in the field of view (FoV) of the image sensor to not having enough images for training deep learning techniques. Practical implementation of indoor positioning techniques needs to leverage the ubiquity of smartphones, which is the case with VLP using complementary metal oxide semiconductor (CMOS) sensors. Images for VLP can be gathered only after the lights in question have been installed making it a cumbersome process. These limitations are addressed in the proposed technique, which uses simulated data of a single LED to train machine learning models and test them on actual images captured from a similar experimental setup. Such testing produced mean three-dimensional (3D) positioning error of 2.88 centimeters while training with real images achieves accuracy of less than one centimeter compared to 6.26 centimeters of the closest competitor.

## I. Introduction

The indoor positioning systems (IPS) have been researched extensively both commercially and in academia owing to the wide array of applications it caters to. While there are several extant positioning techniques, VLP has a unique set of advantages, which makes it viable for further study. The indoor positioning problem consists of two steps, identifying the location of the LED and estimating the receiver location with respect to the LED. Radio fingerprinting[1] and optical camera communication(OCC)[2] have been used to solve the first part.

Several techniques have been used to estimate receiver location using VLP, of which most still use geometric processing and computer vision for localization. A single LED positioning system for circular LEDs was proposed in [3] and a similar computer vision technique was proposed in [4] for rectangular LEDs but both techniques fail when the shape of the LEDs change. While machine learning has been used for receiver tilt correction[5] and regression neural networks have been used for positioning[6] both techniques fail to provide for data augmentation and require cumbersome geometric processing for feature extraction.

This Research is supported by the RIE2020 Industry Alignment Fund - Industry Collaboration Projects Funding Initiative (Award No. I1801E0020) administered by the Agency for Science, Technology and Research (A*STAR), as well as cash and in-kind contribution from Surbana Jurong Pte Ltd.

The use of simulation for data augmentation has been explored in [7], but they fail to take the transmitter details such as luminous intensity and receiver details such as exposure into consideration and end up with a 2D shape projection.

This work proposes a single LED VLP technique using simple feature extraction to employ tree-based machine learning techniques. The dearth of data for training and cumbersome data collection was addressed through simulation, which can also be used for other deep learning models. The proposed technique was shown to outperform standard computer vision and neural network based models.

## II. Methodology

### A. Proposed Structure

The proposed structure outlined in Fig. 1, shows the two major parts of the technique, offline and online process. The first step of the proposed structure is the image simulation using Blender[8], where from features are extracted. Feature extraction is used to convert the image into a list of points, which become the input features of the machine learning model. This simple step removes the need for deep learning models which perform advanced feature extraction from unstructured data such as images. The tree-based machine learning models are then trained using the list of points as input and the 3D location as the output. The trained machine learning model is then used to test performance on the real images captured from a smartphone camera. This produces the location of the light in the receiver coordinate system (RCS).

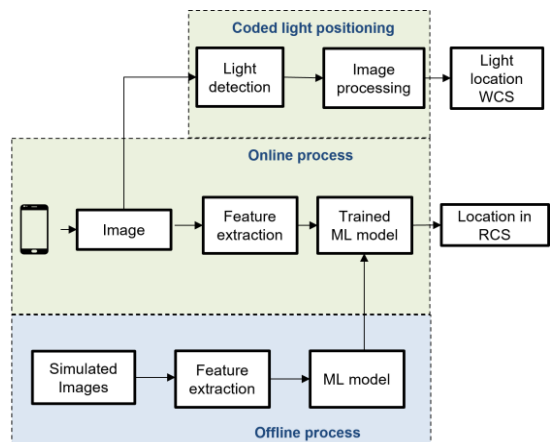

Fig. 1. Overall flow of proposed structure

This however is only a part of the entire process, since the location of the light is needed to identify the receiver location in the world coordinate system (WCS). This is achieved using the high switching rate of LEDs. A unique ID is assigned to an LED and the ID is encoded using differential manchester encoding and beamed to the receiver using on off keying (OOK) and due to the rolling shutter effect of CMOS sensors, a temporal record of the different states of the transmitter are captured in a single image. This is then decoded using the technique proposed in [2]. The focus of this work is on 3D location estimation of the receiver with respect to the transmitter since the demodulation technique produces hundred percent detection over the range tested.

*B. Experimental setup*

The experimental setup for data collection to train and test the proposed technique is shown in Fig. 2(a), where a grid is made on the ground using tape covering 2m by 2m with each line in the grid, both horizontal and vertical being spaced 20 cm apart. This grid will act as a reference for accurate data collection using smartphones since it is done by placing the camera on the tripod with the screen facing the light. The light is 256 cm from the ground and by controlling the height of the tripod the distance from the light is controlled. The images were collected for a 1.2 m by 1.2 m grid at four different heights, 1.23 m, 1.3 m, 1.6 m and 1.66 m away from the transmitter. Here again ten images were collected at each of the 49 locations for both heights with the device orientation and tilt being changed randomly for all images to provide a wide dataset for testing generalization of trained networks.

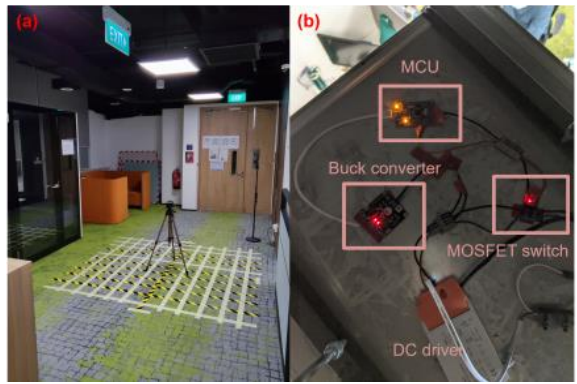

Fig. 2. (a) Experimental setup (b) transmitter components

The components used to transmit the ID using the LED are shown in Fig. 2(b), where the STC12C5A60S2 board was used as the micro controller unit(MCU) which encodes the ID and sends the signal to the n-channel MOSFET, which turns on or off the supply from the DC driver to the LED based on the input signal. A buck converter was used to step down the LED supply to power the MCU. Since manual control of the exposure settings was required an Android application was developed to capture images as shown in Fig. 3(a), where the gray scale image of the LED transmitting an ID is seen in the viewfinder. Owing to the high shutter speed, we can see the clear separation of the light and the background. The features of interest are the corners of the light, which can be extracted using the Shi-Tomasi corner extraction technique[9] In the case of images with the transmitted ID as in Fig. 3(a), the image was dilated to combine the bars which can then produce corners. The parameters of interest in this case are shutter speed which determines the maximum frequency a device can decode, where it is important to note that the shutter speed must be higher than the frequency of operation since the lights are usually at the ceiling at least a couple of meters from the user and the image captured from such distances will have the light cover a small portion of the image. The other parameter of interest is the ISO, which is a measure of the sensitivity of the CMOS sensor to light hitting it. If this number is high, it will pick up low intensity lights which could lead to multipath effects from reflections due to windows or even on walls if this parameter is high enough making it difficult to identify the light bounding box in the image. These values can be modified to suit the problem space using the smartphone application developed as shown in Fig. 3(b). In this study using the Redmi Note 9 Pro front camera the exposure time and ISO take the values of 68 microseconds and 100 respectively throughout for all images unless mentioned otherwise. The plus and minus buttons next to the parameter on the camera settings overlay to be changed can be pressed to change them and the current value is displayed between the buttons.

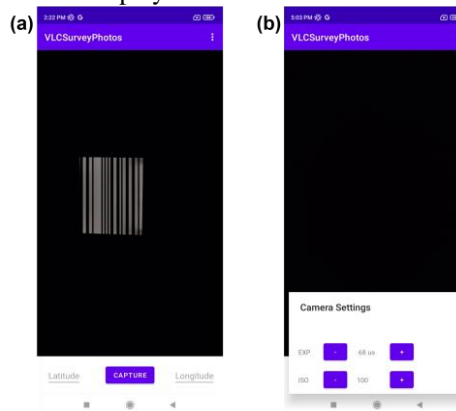

Fig. 3. Receiver Android application

The next step is capturing the images which can be done by clicking the capture button on the bottom of the screen. A sample of an image in the view finder of the application is shown in Fig. 3(a), with the parameters set at the aforementioned exposure time and ISO values which yields a clear image of the strides of an eight-bit long code at ten kilo hertz frequency. The captured image can be named by entering the location coordinates on both sides of the capture button since this also serves the purpose of image collection for training and testing data in the case of intra-cell positioning using a single transmitter. The captured image is then to be processed to decode the location ID being transmitted by the LED, which can then be matched to a database of known location IDs to identify the transmitter location. For each of the grid locations ten images were captured at each height of which two were selected as test data and the remaining eight were used as training data.

*C. Image simulation*

Data augmentation techniques are generally employed in standard deep learning-based classification problems. These range from scaling, rotating to inverting images which in this case would make the image unusable. However, this is one of the challenging applications for data generations since it will be a three-dimensional regression problem eventually when defined as a camera relocalisation problem with a six degree of freedom quaternion as its output. Since collection and labelling of images accurately is time consuming and a seemingly endless amount of data can be collected depending on the accuracy of detection expected simulation using Blender was used to augment data collection. The simulation screen from Blender is shown in Fig. 4, where an area light was modelled to replicate the specifications of the LED used for testing. A 59.5 cm square LED panel from Lite Unite, DWUGR606036 was used for testing producing 3600 lumen with a color temperature of 4000K. The area light was modelled as a plane with an emission shader as shown in the bottom panel, where the color temperature was replicated using a blackbody node with the temperature set to the appropriate value and the polar curve of luminous intensity was used to produce an illuminating engineering society(IES) file to model the throw pattern. The IES node was used to set the appropriate signal strength of the emission shader.

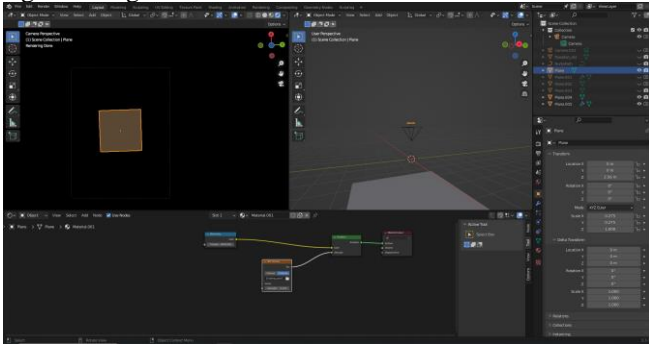

Fig. 4. Blender simulation screen

The image simulated from this technique is shown in the left panel in Fig. 4, where the black background is obtained by setting a low exposure value. Since the shutter speed for VLC and reading the ID from the coded light module must be very high only the brightest parts of the image, which in this case is the light, are seen with all background features being lost. This also ensures that feature extraction from the image becomes much easier owing to the simpler image and also enables reuse of the image for all lights with the same shape since the background features are ignored. However, the pattern formed from images transmitting the ID is not simulated since the corners of the lights are the only features being used for training the machine learning model. The camera used for rendering images in Blender was placed at different positions and at different orientations in the space below the light controlled by location and rotation values. The images were generated with a 3:4 aspect ratio, which is the most common choice for smartphone sensors, at the same resolution, 1728 X 2304, as the test image to ensure compatibility between simulated and test data. The data was simulated at the same grid locations at two of the same heights 1.3 m and 1.66 m from the light with two more heights different from the experimental data capture at 1.76 m and 1.56 m from the transmitter, which will be used to test the generalizability of the trained model. The receiver orientation was swept in complete circles on roll, pitch and yaw values in increments of 45 degrees with only the images where all four corners of the light were in the FoV of the camera were retained, which produced 7982 images for all four heights. Since these images were simulated, the corners were also labelled using ray casting to be the appropriate points corresponding to the LED corners which can be cumbersome in the actual data gathering process. The LED panel used here is a square and without any background features the images will look similar along any of the four sides leading to erroneous results without additional information. The simulation process simplifies this allowing the 3D position to be estimated without pose or orientation information.

III. RESULTS AND DISCUSSION

A. Model selection

The corners were extracted from both the simulated and real datasets and ordered with the list of points ordered in the same sequence manually in the case of real images with simulated images being generated with the corners labelled. The real dataset will be used as the test set in this section. Here, the images at 1.3 m and 1.66 m from the transmitter were split into train and test sets with 2 images at each grid location for the latter and the rest for the former. The test set has 196 images and train set has 784 images in the real image dataset. All the simulated images were used for training, which was 1163 at 1.3 m and 2516 at 1.66 m, totals to 3679 images. The simulated dataset is 4.6 times the real dataset owing to the ease of collecting and labelling data. Three models were trained on the real dataset and tested at 1.66 m from the transmitter. Two tree-based ensemble techniques, random forest[10] and extreme gradient boosting(xgboost)[11] were tested along with a multi layer perceptron[12].

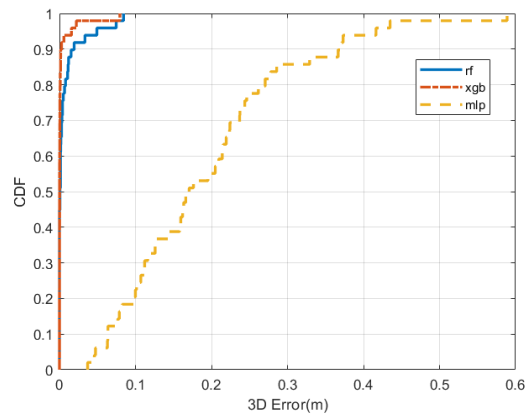

Fig. 5. CDF of 3D positioning error for different models

The 3D positioning error is the Euclidean distance between the estimated location and the actual ground truth. The cumulative distribution function (CDF) of the 3D positioning errors is shown in Fig. 5, where the tree-based techniques outperform the neural network. The models were implemented using the scikit-learn package[13], with default values for all parameters apart from number of estimators for the tree-based techniques, which was changed to 150 and for the neural network five hidden layers with 200 nodes in each were used. The 3D positioning error using the neural network for 90% of points is shown to be less than 40 cms while random forest, which is the worse of the tree-based techniques, has all points less than 10 cms. The marked improvement is expected since the images are converted to a list of points making it a structured dataset. The tree-based techniques are shown to outperform neural networks across multiple structured datasets[14].

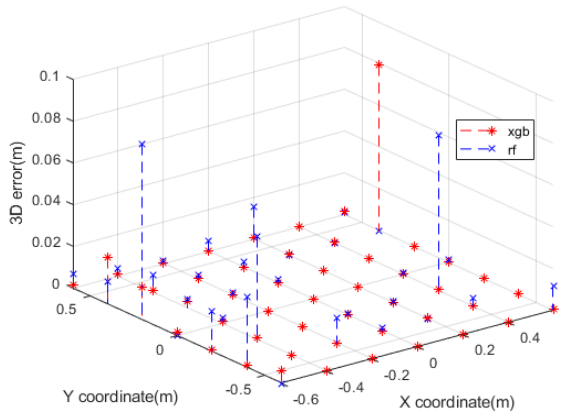

Fig. 6. 3D positioning error for tree-based models

Among these tree-based techniques, the mean 3D positioning error at each grid location is shown in Fig. 6, where the xgboost model is marked with asterisk and the random forest model is marked with cross. The results of the random forest model shows that most of the error comes from outermost points in the grid and some from points closer to the center, with multiple points producing more than 5 cm of error. In the xgboost model all the errors are in the outermost points with only one point producing more than 5 cm error. This explains the better overall performance in the case of the xgboost model over the random forest model. The 3D error CDF also shows that though both models have similar maximum errors, the error for xgboost is lower across all the points in the test dataset. Thus, the xgboost model was chosen for subsequent testing.

*B. Effect of simulated data*

TABLE I. MEAN 3D POSITIONING ERROR RESULTS

| Train | Test | Distance from light (m) | Mean 3D error (cm) |
|---|---|---|---|
| Simulated | Real | 1.66 | 3.11 |
| | | 1.3 | 2.65 |
| | | 1.6 | 5.32 |
| | | 1.23 | 4.9 |
| Real | Real | 1.66 | 0.29 |
| | | 1.3 | 0.104 |
| | | 1.6 | 1.32 |
| | | 1.23 | 1.17 |
| state of the art | | 1.66 | 6.35 |
| | | 1.3 | 6.17 |
| | | 1.6 | 8.13 |
| | | 1.23 | 6.07 |

The simulated dataset was used to train a xgboost model, which was tested on the real dataset. The results of the same are to be compared with a xgboost model trained and tested on real images and the closest competitors technique used on the real test dataset. The closest competitor is marked sota in Fig. 7 to indicate the state of the art (sota) results reported by the same[4]. The sota uses computer vision to identify geometric relations between the four points in the image plane, camera coordinate system and world coordinate system. They also use a photo detector (PD) to identify the location of the light in the world coordinate system.

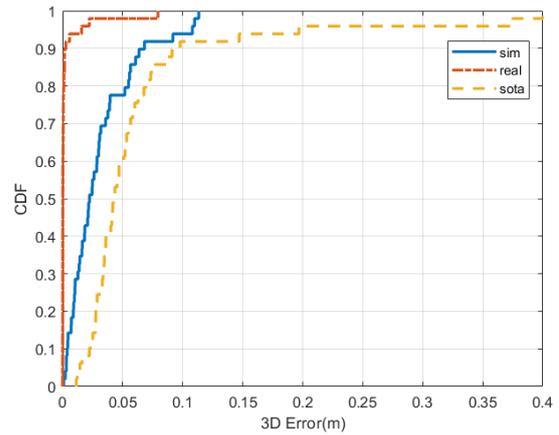

Fig. 7. CDF of 3D positioning error for data at 1.66 m

The model trained on the real dataset produces the best results of the three techniques though it was trained on four times fewer data points as observed from the CDF of 3D error at 1.66 m from the light in Fig. 7. This however, fails to take into consideration the time intensive labelling process of the corner points. The maximum error produced here is less than 10 cm when trained and tested with real images. The maximum error rises to 12 cm in the case of training with simulated images and testing with real images. The sota performs the worst with maximum errors of upto 40 cm. However more than 90% of the points have less than 10 cm of error in all three cases.

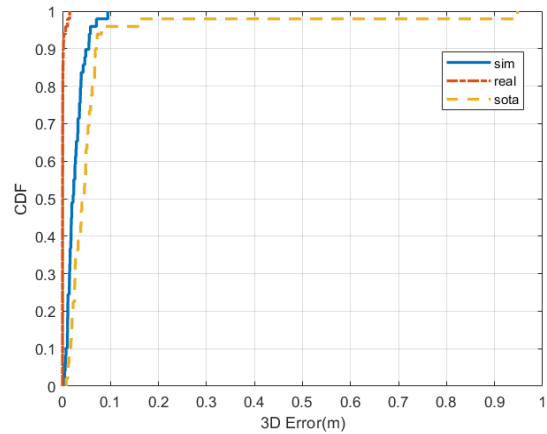

Fig. 8. CDF of 3D positioning error for data at 1.3 m

The CDF of 3D positioning error at 1.3 m from the light is shown in Fig. 8, where the maximum errors for both models trained on simulated and real images decreases but the maximum for the sota increases to 90 cm while the 90% performance improves slightly indicating that there are outlier grid points in the sota affecting the overall performance. The mean 3D positioning errors are listed in table 1, with the train and test columns indicating the training dataset and testing datasets used. The mean errors are consistent with the CDF observed at these two heights, as we move closer to the light the overall positioning error decreases. Though the positioning accuracy achieved in the simulated dataset is lower than the real images, it still is better than the sota by more than 3 cm for both the heights indicating the similarity of the simulated images to real images.

## C. Performance generalization

The results reported thus far have used either simulated or real images from the same heights for training the models. However, this is not a good indicator of the model having learnt the relationship between corner points of the light in the image and the 3D coordinates of the receiver location with respect to the light. In order to test if the model has learnt this relationship two different heights of the real images were used as test datasets at 1.6 m and 1.23 m from the light. These are just 6 cm and 7 cm away from the original training locations, in order to truly test the generalization of performance on the height axis, two more sets of images were simulated at 1.76 m and 1.56 m from the transmitter. Though the test set at 1.6 m is still close to one of the datasets, the 1.23 m test set can be used to gauge consistency of results since it is further from both the datasets. The number of images for this simulated set has changed owing to the change in distance from the light, with 2683 images at 1.76 m and 1620 images at 1.56 m, to 4303 images. The real dataset however was kept at 1.66 m and 1.3 m owing to the difficulty in data collection and labelling. Since the proposed technique involves the use of simulated images rather than real images a new real image training set was not created for the new heights at which images were simulated.

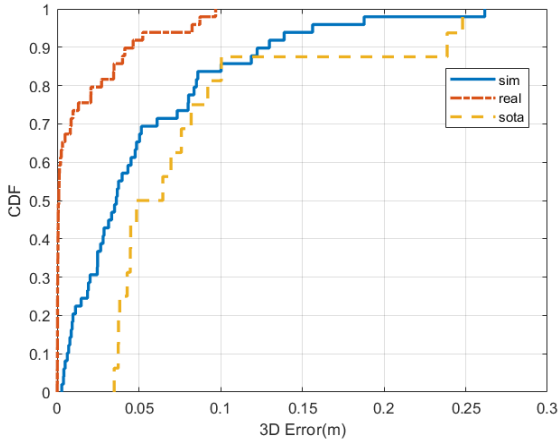

Fig. 9. CDF of 3D positioning error for data at 1.6 m

The CDF of the 3D positioning error for test data at 1.6 m from the light is shown in Fig. 9, where the model trained on real images performs the best with the maximum error still lower than 10 cm. The model trained on simulated data shows a marked decrease in performance owing to the new dataset further away from test points and has a higher maximum error than the sota in this case. However, 90% of the points have less than 15 cm error in the simulated dataset while the sota has the same mark at less than 25 cm. This marked difference in performance is observed owing to the distance from the light being higher for this dataset and the simulated points being further away on average from the test points. The CDF of 3D positioning error in the case of test data at 1.23 m from the transmitter is shown in Fig. 10, where the model trained on the simulated dataset performs better with a marked reduction in the maximum error from 25 cm to less than 18 cm. The sota achieves similar maximum error but more than 90% of the points are observed to have an error of less than 10 cm which once again indicates outliers in the case of sota causing performance issues compared to the simulated results in [4].

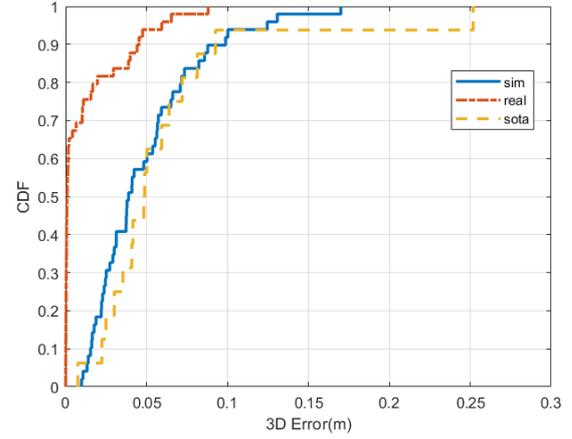

Fig. 10. CDF of 3D positioning error for data at 1.23 m

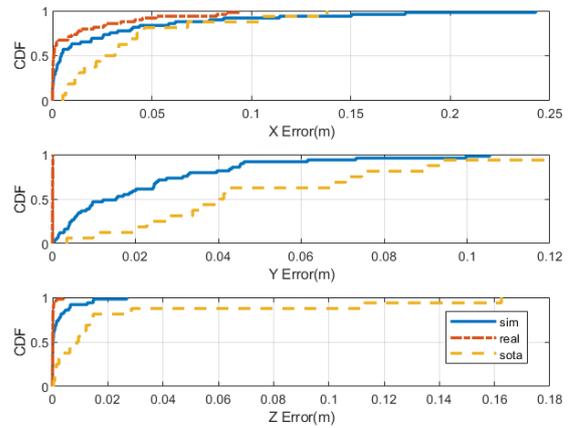

Fig. 11. CDF of three axis errors for data at 1.6 m

The CDF of individual deviations of the estimated locations from the ground truth on all three axes is shown in Fig. 11, where the sota performed worse on all three axes. The simulated data observes a higher error on the x and y axis than the z axis indicating the robustness of the relationship learnt by the model. The x axis produces highest error for all three techniques with only the model trained on real images managing a 90% mark less than 5 cm error. In both the other axes almost no error is observed in the real model, but the simulated model performs better than the sota in all three axes individually producing the lowest error in the z axis. The CDF of individual deviations for the data 1.23 m from the light is shown in Fig. 12, where the z axis error is the lowest for all three models owing to the receiver's proximity to the light. The maximum errors are produced in the x axis once again but this time both sota and the simulated model perform much closer to the real images model on the x and z axis with the y axis producing the highest difference between them. From the table 1, the mean positioning error is also consistent with the observed results thus far, the model trained on real images performs the best across the board for all heights but this data collection strategy is not scalable when applying to deep learning models. The simulated models perform much better than the sota by 3 cm at 1.6 m and 2 cm at 1.23 cm, which apart from the proximity to the transmitter is also driven by anomalies in the test dataset at 1.6 m which causes an increase in error across all three models but the most pronounced errors in sota. This cannot be due to height generalization testing since the sota employs a computer vision based technique and does not rely on data for modelling.

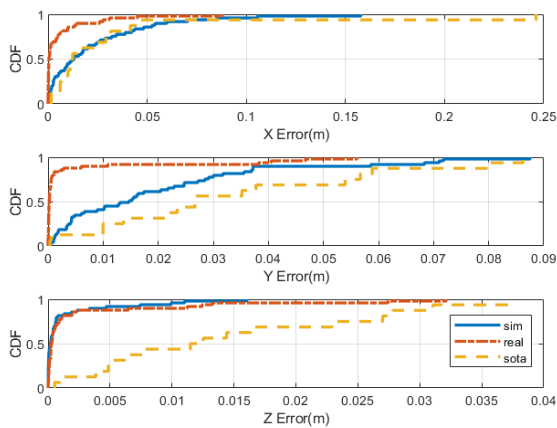

Fig. 12. CDF of three axis errors for data at 1.23 m

## IV. CONCLUSION

We proposed a tree-based VLP technique using simulated data for single LED indoor positioning without the need for data collection and labelling. The model trained on simulated images was shown to perform better than the closest competitor and within 3 cm of mean 3D positioning error from the model trained on real images. The similarity of results obtained between the simulated and real images indicates the photorealism observed in the simulation. The conversion of images to a list of points reduces the unstructured images to structured data enabling the superior performance compared to the closest competitor. The superior performance of tree-based models on structured data is leveraged to obtain these results. The generalization of the models was tested by simulating images further from the test points and the models were shown to perform best on the z-axis with the lowest error among the three axes. The pose estimation and point labelling with sensor fusion will be explored further in subsequent works.

## ACKNOWLEDGMENT

The authors thankfully acknowledge the support from Surbana Jurong-Nanyang Technological University corporate laboratory.